\documentclass[11pt]{article}  
\usepackage{graphicx}
\newcommand{\be}{\begin{equation}}
\newcommand{\ee}{\end{equation}}
\newcommand{\bea}{\begin{eqnarray}}
\newcommand{\eea}{\end{eqnarray}}

\begin{document}
\vspace{.5in} 
\begin{center} 
{\LARGE{\bf A model field theory with $(\psi \ln \psi)^2$ potential: Kinks with super-exponential profiles}}
\end{center} 

\vspace{.3in}
\begin{center} 
{{\bf Pradeep Kumar$^{(1)}$, Avinash Khare$^{(2)}$ and Avadh Saxena$^{(3)}$}}\\ 
{$^{(1)}$Department of Physics, P.O. Box 118408, University of Florida, Gainesville, Florida 32605, USA \\
$^{(2)}$Physics Department, Savitribai Phule Pune University, 
Pune, India 411007 \\ 
$^{(3)}$Theoretical Division and Center for Nonlinear Studies, Los
Alamos National Laboratory, Los Alamos, New Mexico 87545, USA}
\end{center}

\vspace{.9in}
{\bf {Abstract:}}  
We study a (1+1)-dimensional field theory based on $(\psi \ln \psi)^2$ potential.  There are three degenerate minima at $\psi = 0$ and $\psi=\pm1$.  There are novel, asymmetric kink solutions of the form $\psi = \mp\exp (-\exp(\pm x))$ connecting the minima at $\psi = 0$ and $\psi = \mp 1$.   The domains with $\psi = 0$ repel the linear excitations, the waves (e.g. phonons).  Topology restricts the domain sequences and therefore the ordering of the domain walls. Collisions between domain walls are rich for properties such as transmission of kinks and particle conversion, etc. To our knowledge this is the first example of kinks with super-exponential profiles and super-exponential tails.  Finally, we provide a comparison of these results with the $\phi^6$ model and its half-kink solution.

\newpage 
  
\section{Introduction}
Among the (topological) kink-bearing theories \cite{raj}, there are the well-known polynomials, the best known of them being the $\phi^4$ field theory followed by several others as well.  There are also the trigonometric (sine Gordon and Double sine Gordon \cite{mk,sp,ph,scott}) and other transcendental functions.  In each case, they have a potential with degenerate minima giving rise to generally stationary domains. The kinks are the domain walls between these minima.   These are solvable for their analytical profiles.  Often the properties of kink-kink and kink-antikink interactions can be discerned from the properties of the potential.  The surprises sometimes come from a numerical simulation of the collisions, such as the transparency of kink and antikink in a sine-Gordon theory \cite{scott} and particle conversion \cite{mk,sp,ph,scott} in the double sine-Gordon theory. 

In the following, we present a study of a one (space plus one time) dimensional scalar field theory with a potential term of the form $(\psi \ln \psi)^2$.   The provenance \cite{kumar} here is from a theory of higher order phase transitions in the limit of infinite order.   Note that most Gaussian field theories start with a term quadratic in the field as well as in its gradient.  The next terms contain the essential physical specifics e.g. quartic or higher order nonlinearities.  Higher order phase transitions \cite{kumar, farid, kks} contain progressively weaker nonlinearities beyond the quadratic term.  Our model potential represents the {\it minimal nonlinearity} in the form $(\psi \ln \psi)^2$.  In one dimension, it supports an analytically solvable, unexpected kink solution with multiple flavors and unique interactions.  

The objective of this paper is to discuss these solutions, which are kinks in a minimal-nonlinearity field theory described by a $(\psi \ln \psi)^2$ potential.  We begin by noting (see Fig. 1) that there are three degenerate minima in this potential, $\psi = 0$ and $\psi = \pm1$.  The potential is defined for $\psi < 0$ by noting that $\ln \psi = \frac{1}{2} \ln \psi^2$.  The functional form of the (0,1) kink is super-exponential, i.e. of the form $\exp (- \exp (-x))$, which arises in the context of extremal statistics and is known as Gumbel distribution \cite{gumbel}. A stability analysis shows that the kinks are stable, the solid body motion takes place in a manner similar to other kinks, at zero frequency and the corresponding wave function is the gradient of the kink profile.  There is a continuum of propagating states that exist only in the $\psi\ne0$ domain. They are repelled from the $\psi = 0$ domain.  The kink-phonon interaction is described by a Morse-like potential that has an eigenvalue spectrum more akin to the Bargmann potentials \cite{bargmann} of an entirely different context.

Adjacent configurations of kinks and antikinks are subject to topological constraints thus resulting in sequences that are specific.  Interactions are repulsive between the kinks (and super-exponential) and are attractive between a kink-antikink pair (which are either exponential or super-exponential).  

Most of the known kink solutions for a variety of potentials harbor kinks with exponential tails except the $\phi^8$ model \cite{lohe}, which has a kink with a power-law tail. Recently, a whole family of potentials has been discovered in which kinks have a power-law tail \cite{kcs, manton19, ivan, ks}. However, the kink solutions obtained in this paper represent the very first example of a super-exponential tail. 

This paper is organized as follows.  Sec. 2 describes the details of the field theory based on the $(\psi \ln \psi)^2$ potential.  Sec. 3 contains the kink solutions and a discussion of their stability.  Sec. 4 includes the topological considerations that determine the special configurations of infinite sequences.  This section also discusses the nature of various kink-kink and kink-antikink  interactions. Sec. 5 contains a comparison of the results with that of a $\phi^6$ model with a half-kink solution. Finally, Sec. 6 provides a summary of the results and speculations on possible future directions. 

\begin{figure}[h] 
\includegraphics[width= 6.0 in]{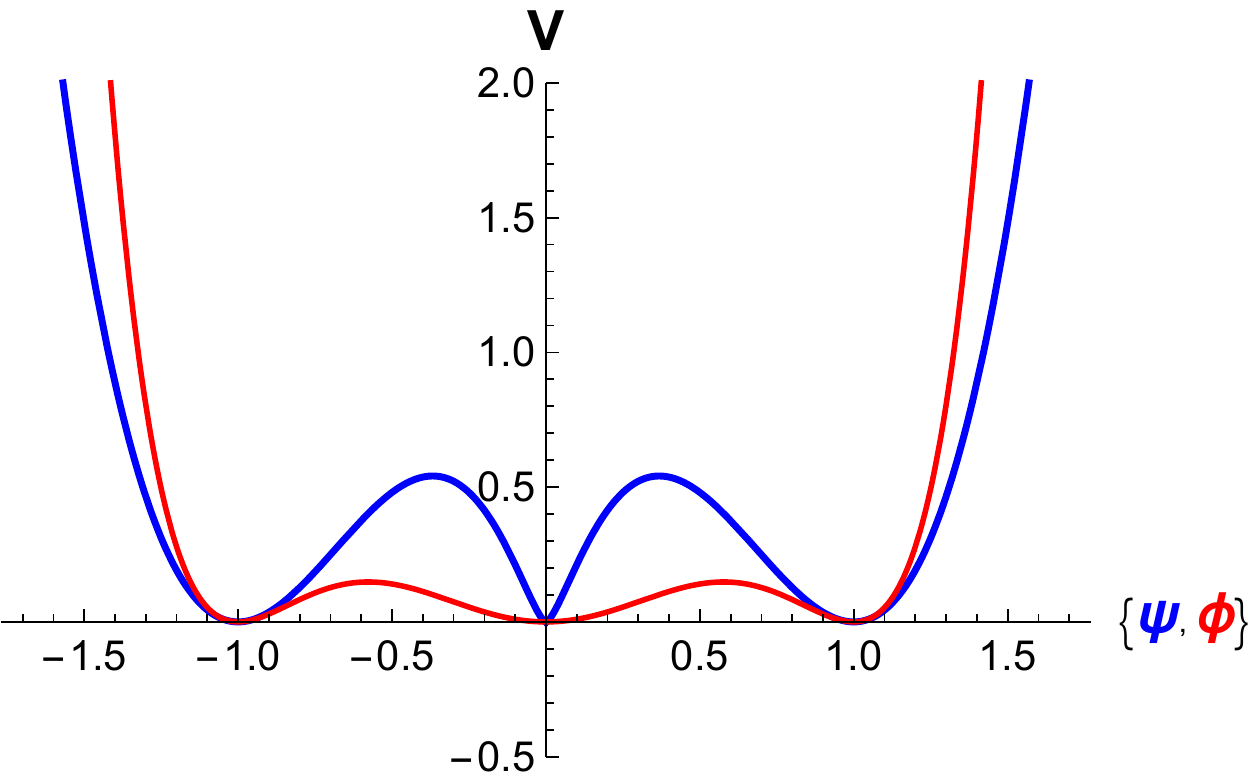}
\caption{Potential $V(\psi)$ associated with the $(\psi \ln \psi)^2$ interaction (blue).   Note that $V(\psi)$ is smooth at $\psi=0$ and there is no cusp.  Comparison with the $\phi^6$ model potential  $V(\phi) = \phi^2(1-\phi^2)^2$ (red).  The conspicuous difference between the two potentials near $V(0)$ leads to super-exponential vs. exponential kink tails. }
\end{figure} 
 
\section{A model field theory}
We consider a field theory described by a free energy functional $F (\psi(x))$ whose minimum provides the Euler-Lagrange equation for the field $\psi(x)$: 
\be 
F = f_0 \int \left[ c|\partial_x\psi|^2 + d (\psi \ln \psi)^2\right] .  
\ee
Here $f_0$ sets the energy scale and $c$ and $d$ are positive, material dependent constants.   As shown in Fig. 1, the potential has three degenerate minima, at $\psi = 0, \pm1$.  Here we use $\ln \psi = \frac{1}{2} \ln \psi^2$, so that negative $\psi$ can be accommodated.  The three minima are separated by two maxima at $\psi = \pm 1/e$. 

We will focus here on a specific dynamics (with $\gamma$ being an effective mobility) described by
\be 
\gamma\frac{\partial^2\psi}{\partial t^2} = - \frac{\partial F}{\partial \psi} . 
\ee 
The frequency dispersion of small amplitude oscillations of wave number $q$  are described by 
\be 
\gamma\omega^2 = V''(\psi_0) + c q^2 . 
\ee 
Here $V' = 2d \psi \ln\psi ( \ln\psi+1)$ is the slope and $V'' = 2d[(\ln\psi)^2 + 3\ln\psi + 1]$ is the curvature of the potential $V(\psi)= d(\psi \ln\psi)^2$.  Note that $V''(1) = 2d = -V''(1/e)$. 

Despite the apparent equivalence of the three degenerate minima, there are differences.  This is first apparent here in that the potential curvature is divergent at $\psi = 0$.  Moreover, there are other related features that appear (and are discussed below) such as a complete blocking of linear excitations, e.g. phonons, from the region $\psi = 0$.  We will defer that discussion until later. 

\begin{figure}[h] 
\includegraphics[width= 6.0 in]{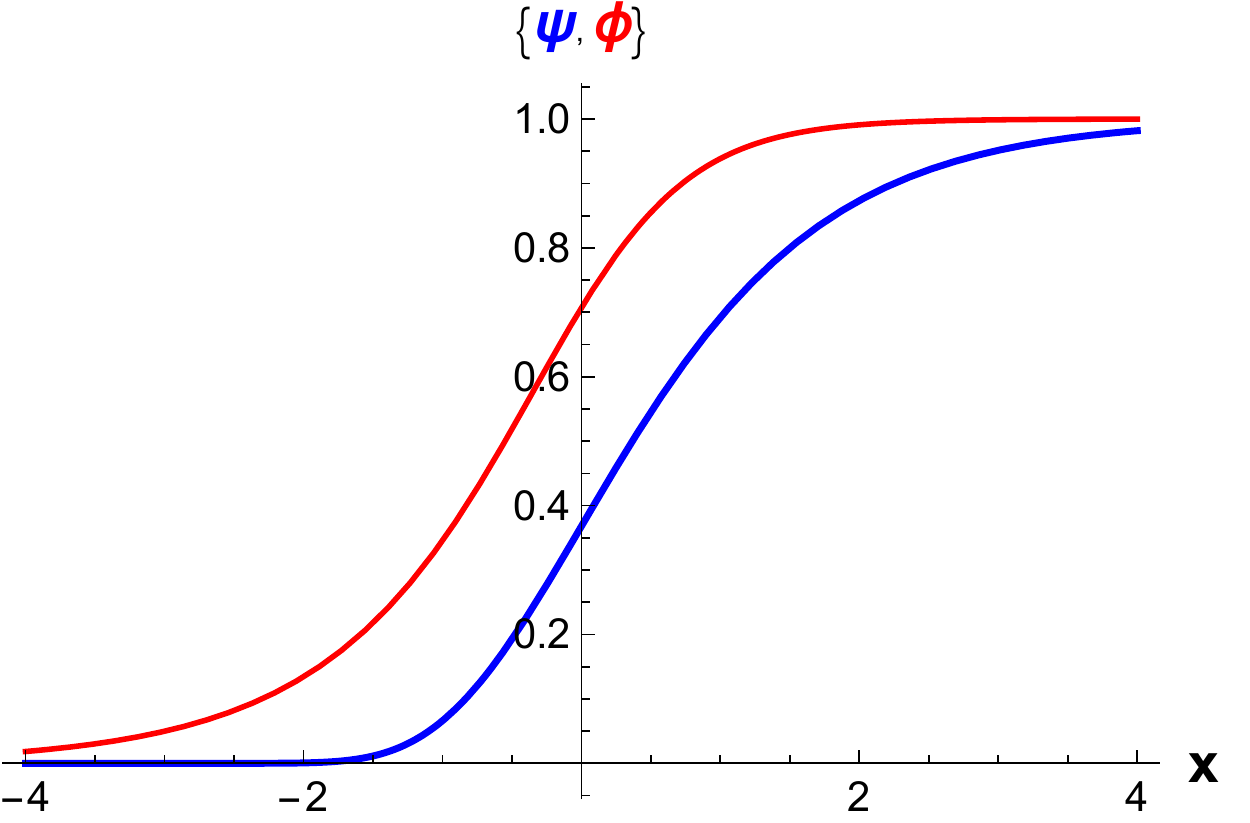}
\caption{Asymmetric kink profile for $\psi(x)$, Eq. (6), with super-exponential $(x < 0)$ and exponential $(x >0)$ asymptotes (blue).  Comparison with the $\phi^6$ asymmetric half-kink, Eq. (11), which has an exponential tail on either side (red). }
\end{figure} 

\section{Kink solutions and their stability}
The Euler-Lagrange equation for a static kink solution becomes 
\be 
-c\partial_{xx}\psi + d\psi\ln\psi(\ln\psi+1) = 0 . 
\ee
The usual practice is to transfer the length scale so that the independent variable is $y = x \sqrt{d/c}$.  The outside energy scale factor then becomes $f_0 \sqrt{cd}$.  After one integration of the equation of motion, we get $\psi_y = \pm \psi \ln \psi$.

{\bf A.} The kink solution connecting $\psi = -1$ at $y = - \infty$ to $\psi = 0$ at $y = \infty$ with the center at $y = 0$ and $\psi(0) = -1/e$ is 
\be 
\psi_A (y) = - \exp (- \exp y).  
\ee
Its energy is $E_A = f_o \sqrt{cd} (1/2\sqrt{2})$.
The antikink solution which has the same energy, a negative slope, connecting $\psi = 0$ at $y = -\infty$ to $\psi = -1$ at $y = \infty$ is $\psi_A (y)' = - \exp (- \exp (-y))$. 

{\bf B.}   The kink solution connecting $\psi = 0$ at $y = - \infty$ to $\psi = 1$ at $y = \infty$ is of the same energy but the profile is given by 
\be 
\psi_B(y) = \exp (-\exp (- y)). 
\ee
The corresponding antikink solution (with the same energy) is given by $\psi_B (y)' = \exp (-\exp(y))$. Note that $\psi_B (y)' = - \psi_A (y)$ and $\psi_A (y)' = - \psi_B(y)$. 

These solutions break the parity symmetry.  This is related to the asymptotic forms of the solution:  For $y  \rightarrow-\infty$, the kink profile for $\psi_A$ is $-1+ \exp(y)$, while for $y  \rightarrow\infty$, it remains a super-exponential.  Between ($-1$, 0), the center at $y =0$ has the value $\psi = - 1/e$. Similarly, for $y \rightarrow\infty$, the kink profile for $\psi_B$ is $1- \exp(-y)$, while for $y \rightarrow-\infty$, it remains a super-exponential. Between (0, 1), the center at $y =0$ has the value $\psi =  1/e$ and the kink is asymmetric (Fig. 2). 

The fluctuations around a kink solution also have novel properties. These are the linear waves (e.g. phonons) interacting with the kinks and are described by a linearized version of Eq. (2).  

Assuming $\delta\psi(y,t) = \psi(y,t) - \psi_0^A = \delta\psi(y) \exp(i\omega t)$, we have, 
\be 
\gamma \omega^2 \delta\psi = -\delta \psi_{yy} + 2(e^{2y} - 3e^y +1) \delta\psi. 
\ee 
This is an eigenvalue problem for a variant of the Morse potential. It contains a repulsive and an attractive term.  The former is square of the latter and the attractive part has a coefficient of 3 (instead of 2 in a regular Morse potential).  This potential has only one bound state at $\omega= 0$, the translational Goldstone mode for the $\psi_A(y)$ kink, which is given by, 
\be 
\Psi_0(y) = e^y e^{-e^y} . 
\ee 
The wave function has no nodes and therefore this must be the lowest eigenvalue.   Since there are no negative eigenvalues here these kinks are therefore stable to emission of linear waves.  There is a continuum of propagating states $(\gamma\omega^2 = 1+ k^2)$, starting with a frequency gap at $\omega = 1$.  These wave functions are perfectly reflected, there is no transmitted component on the {\it right}.  

Similarly, assuming $\delta\psi(y,t) = \psi(y,t) - \psi_0^B = \delta\psi(y) \exp(i\omega t)$, we have, 
\be 
\gamma \omega^2 \delta\psi = -\delta \psi_{yy} + 2(e^{-2y} - 3e^{-y} +1) \delta\psi. 
\ee 
This is also an eigenvalue problem for a variant of the Morse potential (shown in Fig. 3). This potential also has only one bound state at $\omega = 0$, the translational Goldstone mode for the $\psi_B(y)$ kink is given by,
\be 
\Psi_0(y) = e^{-y} e^{-e^{-y}} . 
\ee  
These wave functions are also perfectly reflected, there is no transmitted component on the {\it left}.  The wave function for the bottom of the continuum spectrum has a node at $y = \ln2$.

The super-exponential function, $\psi_B(y) = \exp (-\exp (- y))$, is known in the general area of extreme value statistics and distributions as Gumbel distribution \cite{gumbel}. Presumably, this form arises here due to the field accessing the $\psi=0$ domain with significantly suppressed probability (since $V''(0)$ diverges). 

\begin{figure}[h] 
\includegraphics[width= 6.0 in]{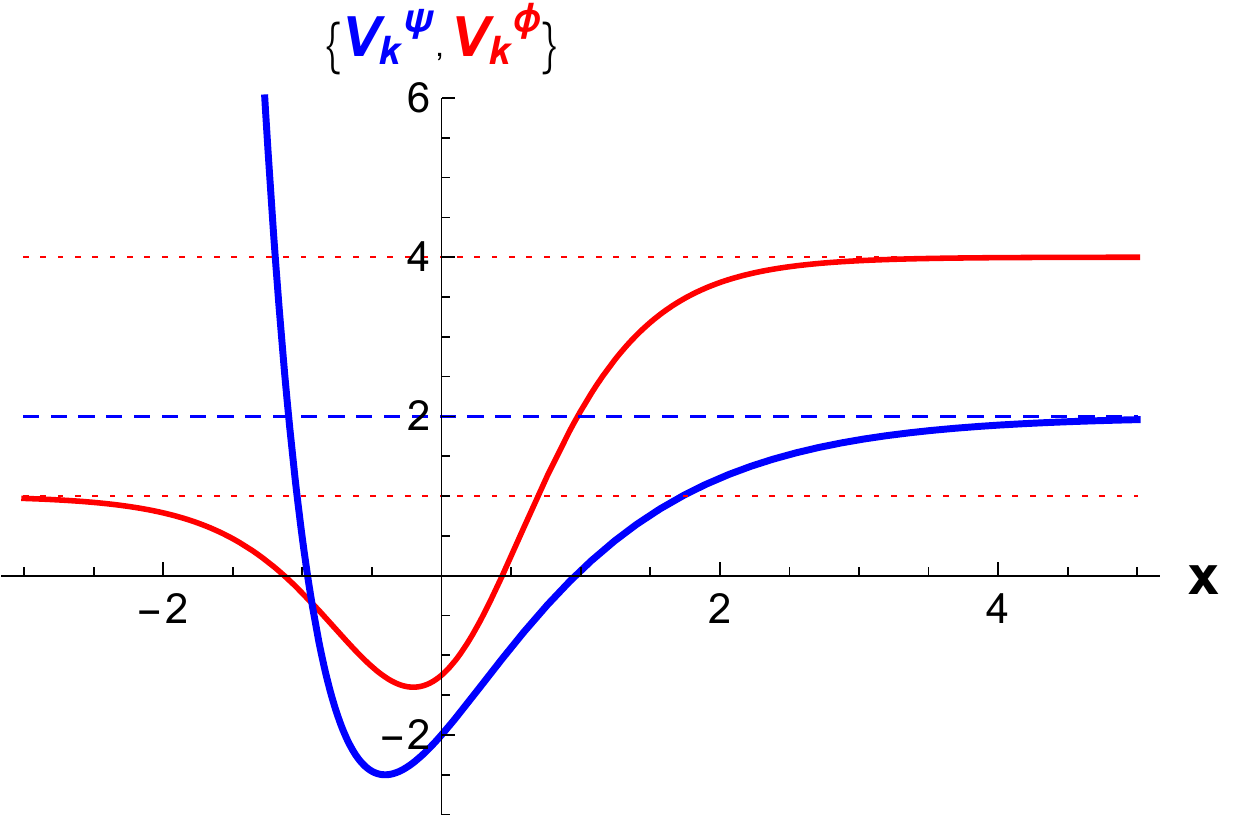}
\caption{Potential associated with phonons (see Eq. (9)) interacting with a type $B$ kink for the logarithmic potential (blue). For $x\rightarrow-\infty$ it diverges whereas for $x\rightarrow\infty$ it asymptotes to 2. Comparison with the potential corresponding to phonons (Eq. (12)) of the half-kink of the $\phi^6$ model (red).  In this case both asymptotes are finite, i.e. 1 and 4, respectively. }
\end{figure}  

\section{K-K' interactions and collisions} 
As we have seen above, this field theory has three degenerate minima (Fig. 1) and therefore that many domains and two types of kinks ($A$ and $B$) subject to topological constraints. Let us consider the interaction between them.  For example, the system may evolve in the state $\psi =1$ in one part and $\psi = -1$ in some different part.  These two domains cannot meet; there must be an intermediate state with $\psi = 0$.  A path drawn from the center of the domain $\psi =1$ and ending at the center of domain $\psi =-1$ will include at least two anti-kinks $B'$ and $A'$.

Figure 4 shows that there are six possible configurations, ($A,$ $A'$) is a bubble of $\psi = 0$ amid $\psi = -1$ on the outside.  ($B$, $B'$) has a bubble of $\psi = 1$ with $\psi  = 0$ on the outside. ($A$, $B$) connects $\psi =-1$ on the left to $\psi =1$ on the right with $\psi =0$ bubble in the middle.  Similarly, ($B'$, $A'$) connects $\psi =1$ on the right to $\psi =-1$ on the left with $\psi=0$ bubble in the middle.  Note that a pair ($A'$, $A$) describes a bubble of $\psi = -1$ in a sea of $\psi = 0$ and ($B'$, $B$) has a bubble of $\psi = 0$ in a sea of $\psi = 1$.

\begin{figure}[h] 
\includegraphics[width= 6.0 in]{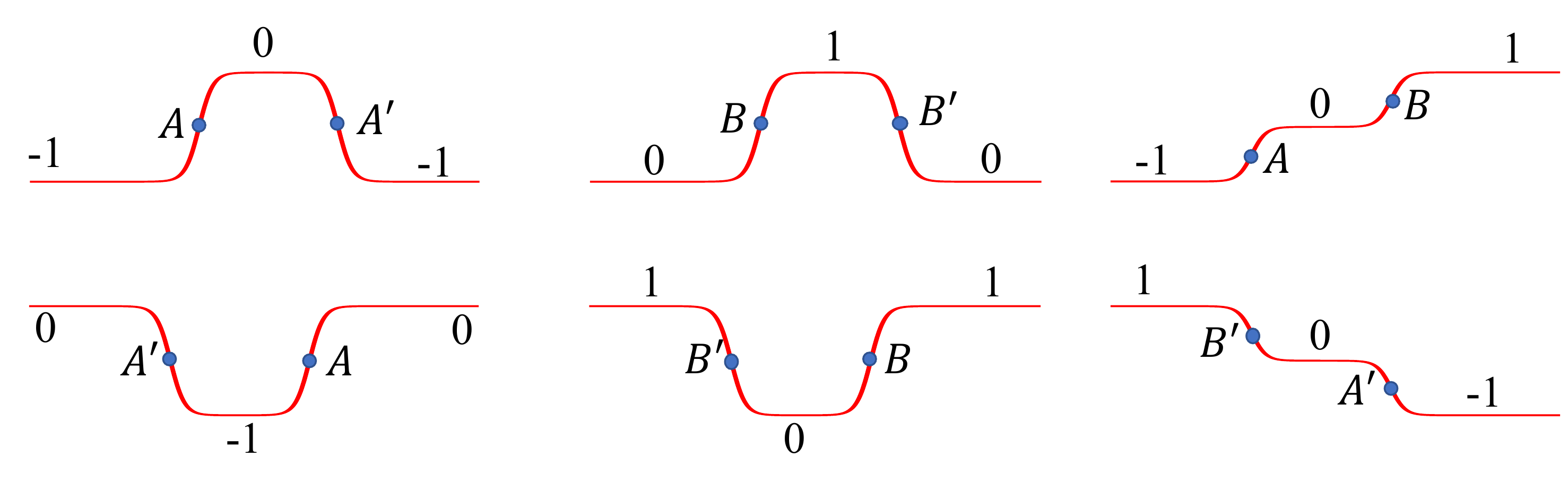}
\caption{Six distinct configurations of various kinks ($A$, $A'$, $B$, $B'$).  Topologically, the $\phi^6$ model kink configurations are the same.}
\end{figure} 

$AB$ kink-kink interaction is repulsive and super-exponential.  On the other hand, $AA'$ kink-antikink as well as $B'B$ (antikink-kink) interactions are attractive and super-exponential. However, $A'A$ and $BB'$ interactions are attractive and exponential. The asymptotic interaction between any of these pairs is inferred from their asymptotic profiles.  Note that the asymptotes are exponential corrections when the kink is approaching $\psi^2 = 1$.  When $\psi= 0$, the approach is super-exponential.  When the overlapping part of the profile is $\psi^2 = 1$, the interaction can be easily calculated and is given by $E_{KK'}=-2 \exp (-d_0)$, where $d_0$ is the kink-antikink separation.

\begin{figure}[h] 
\includegraphics[width= 6.0 in]{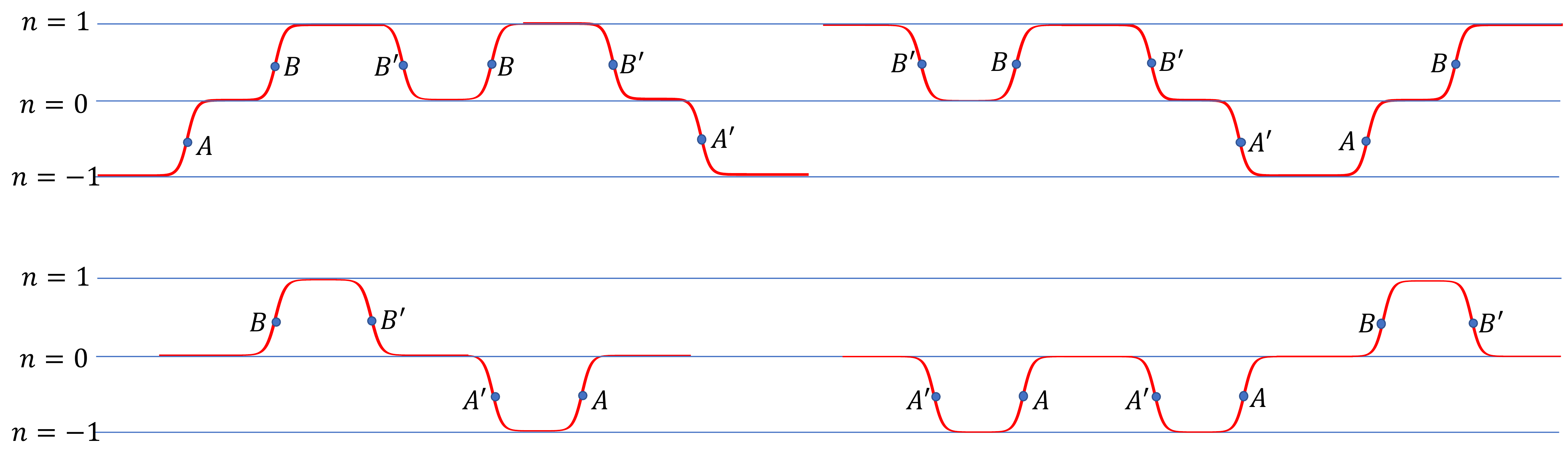}
\caption{Selected representative examples of general kink configurations.  Note that there are many more possibilities than represented here.}
\end{figure} 

There are a limited number of kink-antikink sequences that are possible in an infinite chain (Fig. 5).  These are topological restrictions.  Consider a kink $A$, it must be followed either by an antikink $A'$ or by a kink $B$.  The kink $B$ must be followed by the antikink $B'$.  This is easier seen in a three-rung ladder of $\psi = -1$, $\psi= 0$ and $\psi = 1$.  When the state is $\psi = -1$, it can go to $\psi = 0$ (in kink $A$) and back (in antikink $A'$) an infinite number of times.  That would be an infinite sequence of ($A$, $A'$).  Likewise from the state $\psi = 1$, one can insert an infinite sequence of ($B'$, $B$).  From state $\psi = 0$, one has the choice of antikink $A'$ or kink $B$.

\begin{figure}[h] 
\includegraphics[width= 6.0 in]{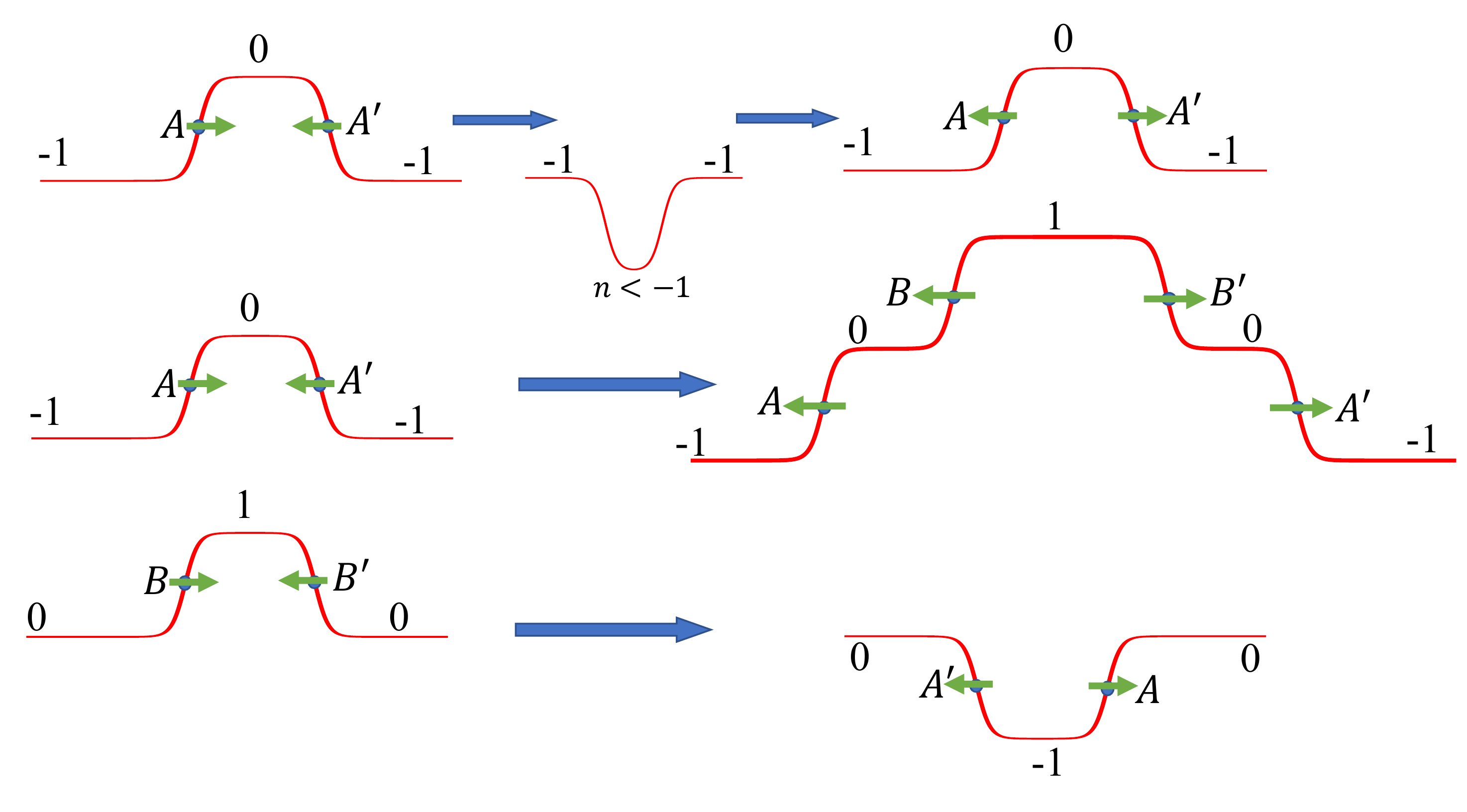}
\caption{Three different collision scenarios depending on the kinetic energy of the colliding kinks. }
\end{figure} 

The kinematics of collisions between allowed pairs of kinks or antikinks has some surprising features (Fig. 6).  Intuitively, it follows that a collision between the pair ($B'$, $B$) refers to a bubble shrinking under the mutual attractive interaction.  The pair may have been launched with some initial kinetic energy. The final state would be the pair reversing their original direction of motion.

Along similar lines, collisions between ($B$, $B'$) or ($A'$, $A$) have conversion properties.  Here the final state can be either the original pair reversing their direction of motion or more likely emerging as an ($A'$, $A$) pair.

\section{Comparison with the $\phi^6$ model} 
A well-known model that has three degenerate minima is the $\phi^6$ model \cite{bk, ss} with the potential $V(\phi) = (1/2) \phi^2(1-\phi^2)^2$ (see Fig. 1).  The minima are at $\phi = 0$, $\pm 1$ and the two local maxima being at $\pm 1/\sqrt{3}$.  The so called half-kink solution \cite{ss} is asymmetric (just like in the current logarithmic potential) and is given by 

\be
\phi(x) = \mp 1 / \sqrt{[1 + \exp(\pm 2x)]}     
 \ee 
with $\phi(+\infty) = 0$, $\phi(-\infty) = -1$ and $\phi(0) = -1/\sqrt{2}$.  For the other kink (preceded by + sign) $\phi(-\infty) = 0$, $\phi(+\infty) = 1$ (see Fig. 2).  As can be easily checked, all the tails have an exponential fall off.  Topologically, the different types of kinks, antikinks and other configurations are similar in the two cases (see Figs. 4, 5, 6), i.e. $\phi^6$ versus logarithmic. 

The analog of Eq. (9) for the stability of the half-kink solution of the $\phi^6$ model in Eq. (11), assuming $\delta\phi(x,t)=\phi(x,t)-\phi_0^K = \delta\phi(x)\exp(i\omega t)$, is 
\be 
\gamma \omega^2 \delta\phi = -\delta \phi_{xx} + \frac{4-10e^{-2x} + e^{-4x}}{(1+e^{-2x})^2} \delta\phi \,, 
\ee 
which is quite different in that in both limits $x\rightarrow\pm\infty$ the kink potential attains a finite value (see Fig. 3).  Nevertheless, just like the logarithmic case (see Eqs. 8, 10), this potential also has only one bound state at $\omega=0$, the translational Goldstone mode for the half-kink, given by 
\be 
\Phi_0(x) = \frac{e^{-2x}}{(1+e^{-2x})^{3/2}} \,. 
\ee 

However, there are some important differences: (i) the asymptotes are different, in the present logarithmic case, one tail is exponential while the other one (near $\psi=0$) is super-exponential, and more importantly (ii) $V''(\psi=0)$ is divergent whereas $V''(\phi=0) = 1$, which is finite. This completely changes how the $\psi=0$ domain behaves in contrast to the $\phi=0$ domain.  (iii) The latter has a well-defined phonon dispersion whereas the former expels phonons (near $\psi=0$).  The exponential tails interact in the usual way, i.e. exponentially \'a l\`a Manton \cite{manton79}.  However, the super-exponential tail leads to a weaker, super-exponential interaction, which is novel.  

There have been significant studies of collision among $\phi^6$ kinks and antikinks \cite{shnir}.  Specifically, as might be expected, results for $B$-$B'$ [i.e. collisions (0,1)+(1,0)] are very different from those of $B'$-$B$ [i.e. (1,0) +(0,1)] collisions. In particular, for $B$-$B'$ collisions, till the velocity $v < v_p$ about 0.289, the kink pair always remained trapped. For $v > v_p$, the collisions yielded a reflected pair of kinks, i.e. $B+B'$ $\rightarrow$ $A'+A$, i.e. (0,1)+(10) $\rightarrow$ (0,$-1$)+($-1$,0). However, $B'$-$B$ collisions revealed a very different picture (as explained in detail in Fig. 1 of ref. [19]). For $v = v_{cr} < 0.0457$ one observes an intricate pattern of escape windows while for $v > v_{cr}$ the kinks have enough energy to always separate. According to ref. [19] this is similar to what happens for collisions in the $\phi^4$ case \cite{campbell}.

Ref. [19] attributes this difference between $B'$-$B$ and $B$-$B'$ collisions to the fact that the stability potential is not symmetrical w.r.t reflections: $x  \rightarrow-x$. So the potential faced by $B$-$B'$ is different than that faced by $B'$-$B$. For small velocities, the adiabatic approximation is valid and using it one can estimate the spectrum of small fluctuations about the potentials experienced by $B$-$B'$ and $B'$-$B$ configurations. From the plots of the potentials \cite{shnir} one sees that the two potentials are different.  Further, the small fluctuation analysis shows that while in the $B$-$B'$ collisions one has only two zero modes, in the $B'$-$B$ case, apart from the zero mode, there are meson states as in the $\phi^4$ case and hence similarities with $\phi^4$ collision case \cite{campbell}. 

The full gamut of possibilities in logarithmic kink collisions are likely similar but will need to be studied numerically beyond what Figs. 4 - 6 suggest.  Because $V''(\psi=0)$ is divergent in the present case, the collision dynamics is expected to be quite different than in the $\phi^6$ case. 

\section{Conclusion}
We have here a field theory based on a potential $(\psi \ln\psi)^2$.  The nonlinear Euler-Lagrange equation can be solved analytically for the kink profile as well as the linear fluctuations eigenvalues and wave functions.  The kinks are novel, super-exponential in analytical form and are described by a Gumbel distribution ($\psi_B$ kink in particular) from extreme value statistics \cite{gumbel}.  The fluctuations are described by the solutions of a Morse-like potential with asymmetric transmission (fluctuations can only survive in the $\psi^2 = 1$ domain).
A comparison with the $\phi^6$ model and its attendant kinks as well as phonons provides useful insights. 

The possible kink-antikink pair sequences have topological constraints. These sequences can alternatively be imagined as bubbles in an onion dome structure.  A domain with $\psi = 0$ may be surrounded by $\psi = 1$ (or $\psi=-1$), which can be surrounded by a larger $\psi = 0$ again.  Inside an $\psi= -1$ (or $\psi = 1$) domain there can be bubbles of $\psi = 0$ only.  The inside bubbles may shrink corresponding to a collision.  Only a $\psi= 0$ domain may contain two possible domains $\psi = 1$ or $\psi =-1$.  A dynamic state where a $\psi =0$ domain surrounds a kernel that breathes between $\psi = + 1$ and $\psi = -1$ may therefore be permitted. 

There is the possibility that given sufficient energy of an (A, AÕ) collision, a (B, BÕ) pair may be created and vice versa.  Here the final state would correspond to a lower speed (A, AÕ) reversing its direction of motion followed by a yet slower (B, BÕ) pair.  Likewise, a collision of (BÕ, B) may result in an additional pair of (AÕ, A) beyond some energy threshold and vice versa.  These multi-particle outcomes seem to be forbidden by some unknown constraints in that, while kinematically possible they are missing in all earlier numerical simulations of collisions. 

\section{Acknowledgment}  We acknowledge valuable discussions with YuXuan Wang, Shizeng Lin and Ayhan Duzgun.  A.K. is grateful to Indian National Science Academy (INSA) for 
the award of INSA Scientist position at Savitribai Phule Pune University. This work was supported in part by the U.S. Department of Energy. 


\end{document}